\documentclass[a4paper]{article}
\usepackage{wrapfig}
\usepackage{INTERSPEECH2021}
\usepackage{xcolor}

\newcommand{\etal}{{et~al.}}

\title{Analysis and Tuning of a Voice Assistant System for Dysfluent Speech}
\name{Vikramjit Mitra$^*$, Zifang Huang$^*$, Colin Lea, Lauren Tooley, Sarah Wu, Darren Botten, Ashwini Palekar, Shrinath Thelapurath, Panayiotis Georgiou, Sachin Kajarekar, Jefferey Bigham}
\address{
  Apple, Cupertino, CA, USA}
\email{(vmitra, zhuang7, colin\_lea, ltooley, sarah\_wu, dbotten, apalekar, sthelapurath, georgiou, skajarekar, jbigham)@apple.com}

\begin{document}

\maketitle
\begin{abstract}
  Dysfluencies and variations in speech pronunciation can severely degrade speech recognition performance, and for many individuals with moderate-to-severe speech disorders, voice operated systems do not work. Current speech recognition systems are trained primarily with data from fluent speakers and as a consequence do not generalize well to speech with dysfluencies such as sound or word repetitions, sound prolongations, or audible blocks. The focus of this work is on quantitative analysis of a consumer speech recognition system on individuals who stutter and production-oriented approaches for improving performance for common voice assistant tasks (i.e., "what is the weather?"). At baseline, this system introduces a significant number of insertion and substitution errors resulting in intended speech Word Error Rates (isWER) that are 13.64\% worse (absolute) for individuals with fluency disorders. We show that by simply tuning the decoding parameters in an existing hybrid speech recognition system one can improve isWER by 24\% (relative) for individuals with fluency disorders. 
  Tuning these parameters translates to 3.6\% better domain recognition and 1.7\% better intent recognition relative to the default setup for the 18 study participants across all stuttering severities. 
\end{abstract}

\section{Introduction}
\label{sec:introduction}
Dysfluencies are common artifacts present in conversational speech and are especially prevalent in people who stutter. Frequent occurrences of stuttered events can reduce fluency in speech, which can impact interaction with voice assistants (VA) such as Alexa, Siri, Google and Cortana~\cite{BrewerCSCW18,ClarkCUI20,USAToday,Moolya,Slate}. In this work we focus on speech with the following major dysfluency types, which are indicative of stuttering: blocks, prolongations, sound repetitions, and word/phrase repetitions, and investigate: 
\begin{itemize}
\item How dysfluencies impact speech recognition accuracy. 
\item Approaches to improve speech recognition performance that do not require model re-training. 
\item How improved speech recognition accuracy can improve the voice assistant experience for users who stutter.
\end{itemize}

It is challenging to detect dysfluencies in speech due to variation in dysfluency types, speaker specific attributes, and contextual dependence. For example, the specific word repetition patterns from one speaker may be very different than in others, or, a speaker may stutter more during a work meeting than when having a conversation with a friend. This phenomena is described in the speech pathology literature~\cite{VanRiper1982,SSI,valente2015event,ingham1993time}, where researchers have investigated diagnostic tools, and proposed strategies to mitigate stuttering in speech. In the speech recognition community, most work has focused on improving recognition performance on fluent speech while leaving those with dysfluent speech behind. As a consequence, for many individuals who stutter it is challenging or impossible to interact with common voice assistants.

Work on dysfluency modeling in the speech community has focused primarily on detecting the presence of dysfluencies in speech or building dysfluency-aware speech recognition models. We recently released 27 hours worth of dysfluency annotations for the Stuttering Events in Podcasts (SEP-28k)~\cite{lea2021sep} and FluencyBank~\cite{FluencyBank} datasets.  
This addressed a major bottleneck for research in dysfluent speech processing, where a lack of data has prevented progress towards the development of accurate dysfluency detectors.
% Prior studies focused on detection of dysfluencies and stutter-features from speech. 
Kourkounakis \etal~\cite{kourkounakis2020detecting} used 800 speech clips (53 minutes) with custom annotations to detect dysfluencies from 25 children who stutter using the UCLASS dataset~\cite{UCLASS} and their followup~\cite{kourkounakis2020b} added synthetic data combining dysfluent speech with the LibriSpeech (fluent) dataset. 
Riad \etal~\cite{riad2020identification} performed a similar task using 1429 utterances in the FluencyBank dataset from 22 adults who stutter.
% Bayer \etal~\cite{BayerlTSD2020} collected a 3.5 hour dataset of German speech from 37 speakers, and developed a model for automated stuttering severity assessment. 
% \cite{Mahesha2016} relied on fewer than 50 samples for pre-trimmed dysfluency classification. 
% In ~\cite{lea2021sep} approximately 26 hours of data was used to train a time-convolutional long short term memory (TC-LSTM) network to classify stutter/non-stutter decisions, and also detect stutter-relevant events. 
Das \etal~\cite{DasNSRE2020} describes work on multimodal detection of stuttering events and shows facial action units, extracted from a camera centered on the user's face, can accurately detect when an event is about to happen.  
While we do explore how a stutter detector can be used to analyze speech in our experiments, the focus of this paper is on speech recognition for voice assistant tasks. 

A recent preprint by Mendelev \etal~\cite{Mendelev2020} is most similar to our work. They built an end-to-end speech recognition model using typical speech and speech with dysfluencies and show a 16\% relative improvement in WER on some voice command tasks for users who stutter compared to a baseline without stuttered speech. To contrast, we focus on how to adapt an existing speech model without end-to-end tuning. 
Some work has explored dysfluency-aware language models or lattice rescoring (e.g.,  \cite{alharbi2018lightly,AlharbiWOCCI17,HeemanInterSpeech16}) to explicitly output dysfluencies such as word or part-word repetitions as part of the prediction and was evaluated on speech from children. Our goal is to output the intended speech or final voice command task requested by a user. 

Standard speech recognition systems are typically trained on thousands of hours of data, far beyond what can reasonably be collected from people who stutter. This work investigates less data-hungry approaches by taking an existing trained hybrid ASR model and exploring if there is a set of different decoding configurations that may improve speech recognition performance for dysfluent speech. Training or fine-tuning the ASR model is outside the scope of this work, as that requires more transcribed speech, and such studies can be explored in the future once more data are available. In addition, we characterize how the severity of stuttering impacts error rates in ASR models, what the likely error types seen in stuttered speech are, and what the impact of speech recognition on task completion is.

\section{Data}
\label{sec:data}
The primary data used in this study were obtained from speakers who stutter performing different voice assistant tasks. A total of 18 adult native US English speakers with varying degrees of stuttering participated. The stutter grade distribution of the participants by the Andrew and Harris scale \cite{andrews1964syndrome} is shown in Table \ref{tab:table1}. Grade 1 is mild stuttering (1-5\% words dysfluent), 2 is moderate (6-20\% words dysfluent) and 3 is severe stuttering (21\%-100\% words dysfluent). The data set consists of 1.6k utterances in total, with an average duration of 6 seconds and a total speech content of approximately 3 hours. Participants were asked to make voice assistant requests ranging from weather, music, web search, etc. These queries usually contain at least six English words of content, not counting content-free fillers, such as “uh” and “um”. All participants' speech was recorded using a mobile phone held approximately 25-30 cm from the participant's mouth, and we refer to this dataset as VA-Dysfluent. 

\begin{table}
%\small
\centering
\caption{Speaker distribution in the VA-dysfluent dataset using the Andrew and Harris's (A\&H) scale.}
\label{tab:table1}
\vspace{1mm}
  \begin{tabular}{|c|c|c|c|}
    \hline
     A\&H scale & Grade 1 & Grade 2 & Grade 3 \\
    \hline     
     \textit{Speakers} & 6 & 7 & 5\\
     \textit{Male} & 5 & 4 & 3\\
     \textit{Female} & 1 & 3 & 2\\
    \hline
  \end{tabular}
\end{table}
\vspace{0.1mm}

A dataset of fluent users making similar voice assistant queries was used as a control for comparison. This dataset, referred to as VA-Fluent, contains approximately 2.7k utterances all made by US English speakers.  
A second dysfluent speech dataset, FluencyBank~\cite{FluencyBank}, was also used for comparison and contains interview recordings from 32 adults who stutter. This dataset was semi-automatically segmented with an average duration of 6 seconds per recording and a total volume of approximately 4.3 hours.

Each dataset was manually transcribed, where two to three transcribers were used to generate the final transcription. The transcription process is curated as follows: (a) initial annotation performed by the transcribers, (b) review the transcription after (a) is completed (c) spot check the transcribed data by an independent reviewer for quality control, and if the transcription fails to meet more than 90\% approval from the quality check, then reiterate through (a). 

There is a discrepancy between what an individual actually says -- dysfluencies and all -- and what they intended to say. This makes annotating this data challenging so we used the following protocol to reduce ambiguity. Given the intended voice assistant query as context, annotators were instructed to transcribe the participants’ intended speech, rather than exact orthographic representation of what was spoken. For example, if the participant said \emph{“who-who was the first per-per-person to walk on the moon”}, then the transcription would be \emph{“who was the first person to walk on the moon”}, as that is likely what the user intended to say. This work uses the intended speech transcription as the ground-truth transcription for all tasks.

\section{Speech Recognition Setup}
\label{sec:speechRecognition}
The Automatic Speech Recognition (ASR) system used in this study is a hybrid deep neural network architecture. The ASR system consists of an acoustic model (AM) that is a deep convolutional neural network, a n-gram language model (LM) with Good-Turing smoothing in the first pass, and the same LM interpolated with a Feed-Forward Neural Network LM in the second pass \cite{gondala2021error}. A set of ASR decoders can be tuned for different purposes such as for performing general dictation tasks or for recognizing assistant query tasks. These decoders are parameterized with terms such as the pruning threshold, beam width, AM weight, word insertion penalty and etc.

\section{Stuttered Speech Recognition Analysis}
\label{sec:analysis}

\subsection{Speech Recognition Analysis}
\label{subsec:ASRanalysis}
We first analyzed intended speech word error rates (isWER) for each dataset by comparing the ASR output with the intended ground truth transcription. For VA query data, both the fluent and dysfluent speech data, a decoder tuned for ``assistant'' mode was used; while for FluencyBank the ``dictation'' model was used as the speech in conversations are more dictation-like. Details about the hybrid AM-LM architecture used in this study can be found in \cite{huang2020sndcnn}. The default ASR ``assistant'' system gave isWER of 5.65\% for fluent speech data and 19.29\% for the stuttered speech data. The ASR ``dictation'' model gave an isWER of 38.66\% for the FluencyBank dataset.

\begin{table}
% \small
\centering
\caption{ASR intended speech error rates for each voice dataset}
\label{tab:table2}
  \begin{tabular}{|c|c|c|}
    \hline
     Error Rates  & VA-Fluent & VA-Dysfluent \\
    \hline     
     \textit{isWER} & 5.65 & 19.29 \\
     \textit{Insertion Rate} & 1.02 & 13.51\\
     \textit{Deletion Rate} & 1.34 & 1.18 \\
     \textit{Substitution Rate} & 3.28 & 4.51 \\
    \hline
  \end{tabular}
\end{table}
\vspace{0.1mm}

Table \ref{tab:table2} summarizes the error rates obtained from queries in VA-Fluent and VA-Dysfluent. Dysfluent speech has more than 3 times the isWER compared to fluent speech and has a very different distribution of error types. 
There is a significant increase ($>$13 times) in insertion rates and a slight increase in substitution rates for dysfluent speech.  Insertion errors are the least common type for fluent speech but the most common for dysfluent speech. This can likely be explained by characterizing specific dysfluency event types. Common dysfluencies including sound and syllable repetitions, sound prolongations, and blocks often lead to inserting partial words and/or miss-recognized words in an utterance. Furthermore, word and phrase repetitions as well as filler words like ``um'' and ``uh'' also increase word insertions.

\begin{table}
% \small
\centering
\caption{Intended Speech Word Error Rate (isWER) by stuttering severity on VA-Dyfluent as graded by a speech pathologist.}
\label{tab:table3}
  \begin{tabular}{|c|c|c|c|}
    \hline
     Stutter Severity  & isWER\\
    \hline 
     \textit{Grade 1} & 8.39 \\
     \textit{Grade 2} & 16.64 \\
     \textit{Grade 3} & 47.86 \\
    \hline
  \end{tabular}
\end{table}
\vspace{0.1mm}

\begin{figure}
\centering
    \includegraphics[width=0.85\linewidth]{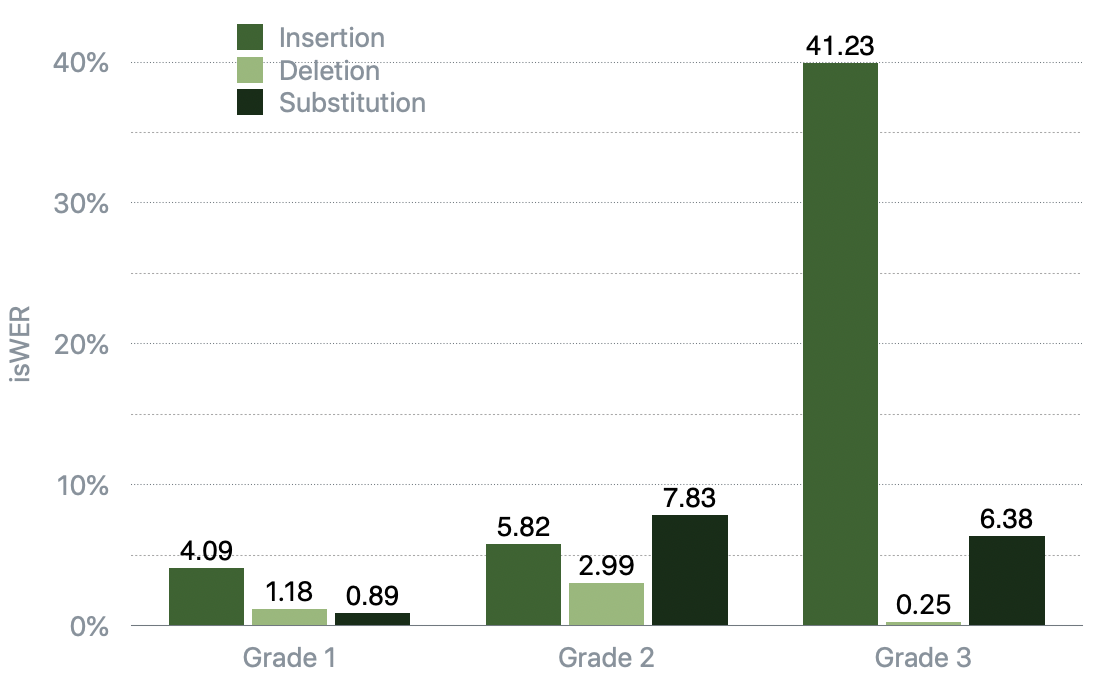}
\caption{Error Rates by A\&H Scale}
\label{fig:error_rate_by_stutter_model_score}
\label{fig:error_rate_by_severity}
\end{figure}

Table \ref{tab:table3} shows a breakdown of error rates for VA-Dysfluent by stuttering severity. As expected, error rates increase substantially as the severity increases. Speakers who stutter severely (Grade 3) have isWER as high as 47.86\%, indicating that the speech recognition system barely works for this group of people. Figure \ref{fig:error_rate_by_severity} highlights the distribution of error rates broken down by stutter severity. 

We examined dysfluent events in subjects who are in Grade 3. Among the 5 subjects, almost all subjects produced frequent sound repetitions. This often occurred at the beginning of an utterance, and in particular was noted to be repeated on phones such as /t/, /d/, /n/, /r/, /s/. One subject produced sound repetitions for duration of up to 15 seconds on initial words in commands, hence contributing severely to the insertion error rate. Repeated sounds /m/, /n/, /l/, and /r/ mid-word were found in some subjects' speech. This, as well as audible and inaudible blocks mid-word, led to many insertion and substitution errors. There are many mitigation strategies meant to reduce stuttering that also contributed to errors in ASR transcription. For example, one user would phonate "ah" before most words, while another user would revise and restart a phrase to avoid or pull-out from stuttering. These patterns contribute to high insertion rates.

\vspace{0.1mm}
\setlength{\textfloatsep}{8pt}

In \cite{lea2021sep} we presented a dysfluency detection model that takes in a sequence of audio and outputs per-frame predictions for whether a sound repetition, block or other dysfluency is occurring. That model is a Convolutional LSTM-based network trained on about 25 hours of dysfluency data from the SEP-28k dataset. 
While that model was not trained on any data from VA-Dysfluent it appears to generalize reasonably well based on qualitative assessment. 
In this work, we used that detection model to estimate the stuttering severity level for each clip in attempts to correlate that with each participant's stuttering grade.
For our approach, we take frame-level predictions for each utterance and aggregate to a range of [0, 100].
Clinical stuttering scores are often based on the count of stuttering events within a given phrase, so taking an average over frame-level dysfluency predictions is a reasonable proxy. 
We classify scores from 0 to 10 as mild, 10 to 30 as moderate, and above 30 as severe. A score of 30 means 30\% of a clip consisted of stuttering events. 

Table \ref{tab:table4} shows the average isWER for all clips that are labeled mild/moderate/severe based on our detection model. 
There is a high correlation between isWERs for both our detection-based model and that of the manually-assigned stutter grades (correlation=0.996).
This means that our model does a good job detecting when we will get a low-quality ASR transcription. 
Unexpectedly, the correlation coefficient between model driven stutter-severity score  and the manually assigned A\&H grade is low: 0.32 with a p-value of $8.46 \cdot 10^{-12}$.
That means that at a per-utterance level our model does not accurately predict the clinical grade. 
While this may seem counter intuitive, it highlights one of the challenges of dysfluency analysis: regardless of stutter severity rating, the same person may have many dyfluencies in one utterance but none in another.

\begin{table}
% \small
\centering
% \caption{ASR isWER by Stutter Model Score}
\caption{Intended Speech Word Error Rate (isWER) by stuttering severity on VA-Dyfluent as computed by the Stuttering Model Score based on~\cite{lea2021sep}.}
\label{tab:table4}
  \begin{tabular}{|c|c|c|c|}
    \hline
     Stutter Model Score  & isWER\\
    \hline 
     \textit{Mild } &  13.64\\
     \textit{Moderate} &  19.42\\
     \textit{Severe} &  33.54\\
    \hline
  \end{tabular}
\end{table}
\vspace{0.1mm}

\subsection{Decoding Parameter Analysis}
\label{subsec:decodeanalysis}

We investigated the impact of ASR decoding parameters on speech recognition performance for dysfluent speech as measured by isWER. Specifically we split each dataset to non-overlapping development (dev) and test sets. We used $\approx80\%$ data as a dev set to tune the ASR system, examining the pruning threshold, beam width, lattice beam, nbest hypotheses, AM weight, word insertion penalty etc., and used the remaining data for testing. 
We also evaluated the model performance with new user testing, where a separate dev and test task was performed. We cross validated leaving one user from each stutter severity grade out for testing (3 subjects) and used the remaining dev set (15 subjects) for learning the decoding parameters. 
In both experiments, two decoding parameters showed a significant impact on isWER for dysfluent (stuttered) speech: AM weight and the insertion penalty. Table \ref{tab:table5} shows how AM weight and insertion penalty variation can cause reduction of isWER for stuttered speech (other parameters are tuned and kept the same). For stuttered speech, the default AM weight and insertion penalty was sub-optimal. An AM weight of 1/30 (in the first scoring and second pass re-scoring) and an insertion penalty of 1, gave the best isWER, which is 24.42\% relatively lower than the default configuration. Parameters including beam size, lattice size and nbest hypothesis also played roles in the performance for dysfluent speech. 

\begin{table}
%\small
\centering
\caption{ASR Error Rates (isWER) of stuttered speech using the  default and optimized AM weight and insertion penalty.}
\label{tab:table5}
\vspace{1mm}
  \begin{tabular}{|c|c|c|c|c|}
    \hline
     Type & Config. & AM & Insertion & isWER  \\
     & & Weight & Penalty & \\
    \hline     
     \textit{VA-Fluent} & Default & 1/15 & 0 & 5.65\\
    \hline
     \textit{VA-Dysfluent} & Default & 1/15 & 0 & 19.29\\
     \textit{VA-Dysfluent} & Tuned & 1/15 & 1 & 18.82\\
     \textit{VA-Dysfluent} & Tuned & 1/15 & 2 & 19.05\\
     \hline
     \textit{VA-Dysfluent} & Tuned & 1/30 & 0 & 15.0\\
     \textit{VA-Dysfluent} & Tuned & 1/30 & 1 & \textbf{14.58}\\
    \textit{VA-Dysfluent} & Tuned & 1/30 & 2 & 14.94\\
    \hline
  \end{tabular}
\end{table}
\vspace{0.1mm}
\setlength{\textfloatsep}{8pt}

For voice assistant tasks, there are a reasonably small number of commands that people are more likely to make (e.g., ``what is the weather''), and therefore it is reasonable to bias an ASR system towards these standard queries. 
This is commonly done by increasing the weight of the language model (or decreasing the AM by proxy) so that the likelihood of recognizing a common command or phrase is high.
In an extreme case, if the LM weight is high, the model will only output phrases it has seen before. We can use this trade-off to our advantage, knowing that dysfluencies such as sound repetitions or prolongations tend to cause errors in the AM part of the system. A low LM weight may retain repetitions whereas a high LM weight is more likely to remove them. As an example, when we applied the default ASR decoder to a dysfluent speech sample it recognized: \emph{“what is my brothers add add add address”}; whereas when we increased the LM weight it more accurately inferred: \emph{“what is my brothers address.”} The multiple sound repetitions “{\textbackslash EY d}” leading to partial word “add” was filtered out. 

\begin{figure}
\centering
\includegraphics[width=0.85\linewidth]{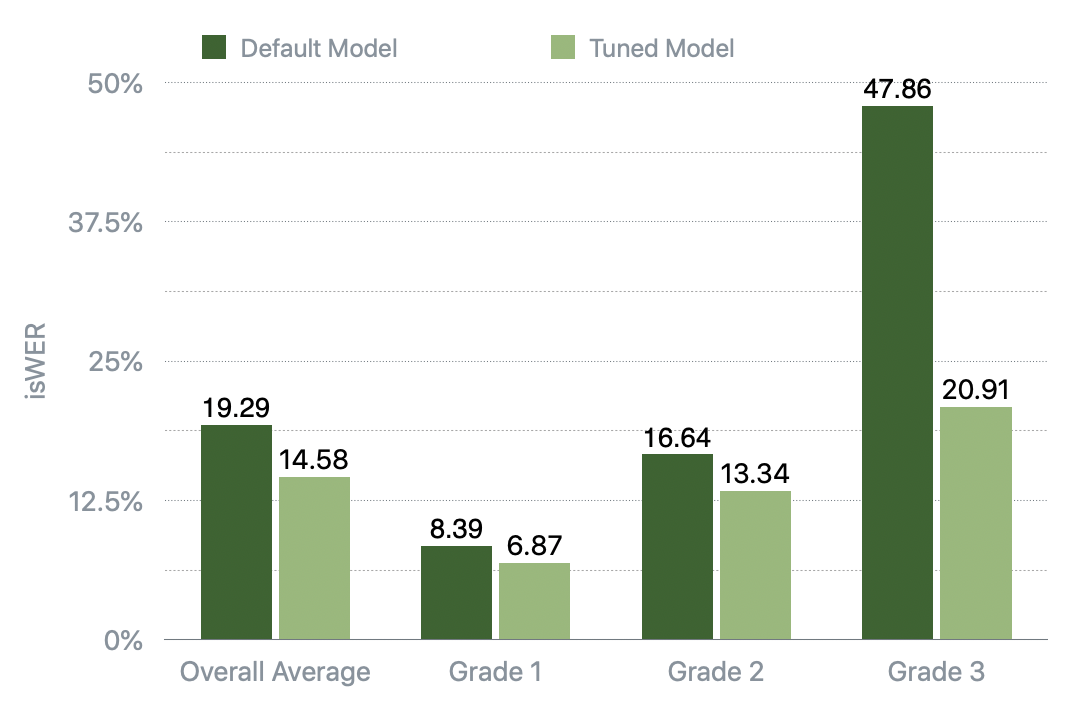}
\caption{ASR Performance Improvement by Stutter Severity between Default and Tuned Models}
\label{fig:WER_between_models}
\end{figure}

The word insertion penalty parameter in the decoder also has a large performance impact by helping to reduce insertion errors. 
When modifying both the LM weight and insertion penalty, the system is able to map speech with dysfluencies to similar and more commonly seen queries. 
For example, the default decoder hypothesized \emph{“when did destiny nearest Walmart near me oh dang,”} whereas the  tuned-ASR was able to predict what the user intended to say: \emph{“when does the nearest Walmart near me open.”}
This decoding configuration led to a remarkable decrease in insertion error rates. 
Figure \ref{fig:WER_between_models} shows detailed results of isWER between the default and the optimal tuned model as a function of stutter severity.

In general, shorter commands tend to be more accurate. When using the tuned model, the average word count when the ASR was correct was 4.6 words, and when the ASR was incorrect it was 6.6 words. For the default ASR decoder the counts are 4.8 and 6.0 respectively. This is intuitive given that for longer phrases there are more opportunities to stutter on a word. We also found that isWER for male speakers were lower than the female speakers and observed significant relative reduction in isWER for both male and female speakers using the tuned decoder compared to the default decoder. 

\vspace{0.1mm}

We did the same decoding parameter tuning for FluencyBank and saw a 5.85\% improvement in isWER from 38.66\% to 36.40\%. 
This minimal improvement is likely due to a domain mismatch where our model is tuned for Voice Assistant tasks and FluencyBank consists of conversational interviews. To quantify the mismatch, we calculated the perplexity of the grader transcriptions on each dataset. Perplexity is 83.98 for VA-Fluent, 88.4 for VA-Dysfluent, and 210.0 for FluencyBank. This indicates that the language model in our system is worse at predicting the phrases in FluencyBank. There is also a mismatch where FluencyBank was collected using different microphones and in noisier environments than the rest of our data. 

\subsection{Domain and intent recognition Analysis}
\label{subsec:domainIntentAnalysis}
Ultimately, for Voice Assistant tasks, our goal is to develop a system that can correctly understand and respond to users queries. WER is the standard metric for measuring ASR errors, yet developing a high-performing VA system necessitates correct understanding of the user's intent, so that the VA can respond to the user's query successfully. 
In this section we investigate VA task performance by measuring accuracy on domain recognition and intent recognition. 
In a typical VA workflow~\cite{chen2019active}, input speech is processed by an ASR to generate text, which is fed as input to the domain classifier to recover the domain of application (phone call, web search, etc.) and then to identify the user’s intent. Intent is used to identify the final response to a user query. If an intent is wrong, then the VA will likely respond with a wrong answer or gives no response (“Sorry, I do not understand.”), leading to a poor user experience.

Table \ref{tab:table5} shows the improvement our tuned ASR system has on VA-Dysfluent using recent domain~\cite{chen2019active} and intent recognition~\cite{ranantha2020intent} models. Overall, there is a 3.6\% and 1.7\% relative increase in domain and intent recognition performance, respectively (statistically significant), obtained from the text generated by the tuned ASR decoder as opposed to the default. On average the tuned system improves domain classification accuracy from 88.05\%  to 91.30\%, consequently improving the identification of the user’s intent. 
Our VA model has a ``garbage'' intent for commands that it does not understand. The tuned model reduces garbage intent predictions by 39.02\%, meaning it is much more likely to provide a helpful response to a user instead of an error message. 
While there continues to be a gap in intent performance between mild and severe stutters, our tuned system demonstrates a 10.6\% improvement in domain accuracy and 7.88\% in intent accuracy for severe stuttering over the baseline. 

\begin{table}
%\small
\centering
\caption{Domain and intent recognition accuracy of stuttered speech using default and optimized decoding parameters.}
\label{tab:table6}
\vspace{1mm}
  \begin{tabular}{|c|c|c|c|}
    \hline
    Severity & Model & Domain Acc. ($\%$) & Intent Acc. ($\%$)  \\
    \hline     
    Grade 1 & Default & 95.25 & 95.06\\
    Grade 1 & Tuned & 95.82 & 95.31\\
    \hline
    Grade 2 & Default & 89.96 & 90.83\\
    Grade 2 & Tuned & 91.15 & 91.28\\
    \hline
    Grade 3 & Default & 76.79 & 78.01\\
    Grade 3 & Tuned & 87.40 & 85.89\\
    \hline
  \end{tabular}
\end{table}
\vspace{0.1mm}

\section{Conclusions}

In this work we quantified performance of speech recognition systems on individuals who stutter in efforts to improve voice assistant systems for everyone. 
We characterized performance of an existing ASR system and identified two dominant error types: insertions and substitutions. 
We found many errors were caused by ambiguity in the acoustic model and observed improvements by increasing the insertion penalty and reducing the acoustic model weight -- thereby increasing reliance on the language model. 
While seemingly simple, changing these decoding parameters resulted in a 24\% relative reduction in isWER. 
We also observed improvements in domain and intent recognition on stuttered speech assistant queries as a consequence of improved speech recognition performance, especially for severe stuttering. 
While perhaps not as technically innovative as some other work in this area, simple changes to existing decoders have the potential to materially improve voice assistant interactions for individuals who stutter. 

Future studies should work to increase the number of participants to further validate these models as there is significant variation in dysfluency patterns across all people who stutter. Existing data is trained on US English and it is unclear how dysfluencies vary across global populations so gathering speech from other language and dialects will be important. 
% In the future we hope to contribute to larger dysfluent speech datasets with transcription will facilitate improving the overall voice assistant performance by enabling the training of robust acoustic and language models.
Future technical studies may work towards model personalization and explore acoustic- or language model adaption. Furthermore, much of the speech community is moving to end-to-end speech recognition models, and it will be important to understand the impact this has on dysfluent speech.

\bibliographystyle{IEEEtran}

\bibliography{interspeech2021vm}

\begin{thebibliography}{10}
\providecommand{\url}[1]{#1}
\csname url@samestyle\endcsname
\providecommand{\newblock}{\relax}
\providecommand{\bibinfo}[2]{#2}
\providecommand{\BIBentrySTDinterwordspacing}{\spaceskip=0pt\relax}
\providecommand{\BIBentryALTinterwordstretchfactor}{4}
\providecommand{\BIBentryALTinterwordspacing}{\spaceskip=\fontdimen2\font plus
\BIBentryALTinterwordstretchfactor\fontdimen3\font minus
  \fontdimen4\font\relax}
\providecommand{\BIBforeignlanguage}[2]{{%
\expandafter\ifx\csname l@#1\endcsname\relax
\typeout{** WARNING: IEEEtran.bst: No hyphenation pattern has been}%
\typeout{** loaded for the language `#1'. Using the pattern for}%
\typeout{** the default language instead.}%
\else
\language=\csname l@#1\endcsname
\fi
#2}}
\providecommand{\BIBdecl}{\relax}
\BIBdecl

\bibitem{BrewerCSCW18}
R.~Brewer, L.~Findlater, J.~Kaye, W.~Lasecki, C.~Munteanu, and A.~Weber,
  ``Accessible voice interfaces,'' in \emph{CSCW}, 2018.

\bibitem{ClarkCUI20}
L.~Clark, B.~Cowan, A.~Roper, S.~Lindsay, and O.~Sheers, ``Speech diversity and
  speech interfaces: Considering an inclusive future through stammering,'' in
  \emph{Conversational User Interfaces}, 2020.

\bibitem{USAToday}
K.~Wheeler, ``For people who stutter, the convenience of voice assistant
  technology remains out of reach,'' USA Today (online), Jan 2020.

\bibitem{Moolya}
P.~Soundararajan, ``Stammering accessibility and testing for voice assistants
  \& devices,'' Personal Blog (online), April 2020.

\bibitem{Slate}
M.~Corcoran, ``When alexa can’t understand you,'' Slate (online), Oct 2018.

\bibitem{VanRiper1982}
C.~Van~Riper, ``The nature of stuttering (2nd ed.),'' \emph{Applied
  Psycholinguistics}, 1983.

\bibitem{SSI}
G.~Riley, ``{SSI-4} stuttering severity instrument fourth edition,''
  \emph{Austin, TX: Pro-Ed}, 2009.

\bibitem{valente2015event}
A.~Valente, L.~Jesus, A.~Hall, and M.~Leahy, ``Event-and interval-based
  measurement of stuttering: a review,'' \emph{IJLCD}, 2015.

\bibitem{ingham1993time}
R.~Ingham, A.~Cordes, and P.~Finn, ``Time-interval measurement of stuttering:
  Systematic replication of {Ingham}, {Cordes}, and {Gow} (1993),''
  \emph{Journal of Speech, Language, and Hearing Research}, 1993.

\bibitem{lea2021sep}
C.~Lea, V.~Mitra, A.~Joshi, S.~Kajarekar, and J.~P. Bigham, ``Sep-28k: A
  dataset for stuttering event detection from podcasts with people who
  stutter,'' \emph{arXiv preprint arXiv:2102.12394}, 2021.

\bibitem{FluencyBank}
N.~Ratner and B.~MacWhinney, ``Fluency bank: A new resource for fluency
  research and practice,'' \emph{Journal of Fluency Disorders}, 2018.

\bibitem{kourkounakis2020detecting}
T.~Kourkounakis, A.~Hajavi, and A.~Etemad, ``Detecting multiple speech
  disfluencies using a deep residual network with bidirectional long short-term
  memory,'' in \emph{ICASSP}.\hskip 1em plus 0.5em minus 0.4em\relax IEEE,
  2020.

\bibitem{UCLASS}
S.~Devis, P.~Howell, and J.~Batrip, ``The {UCLASS} archive of stuttered
  speech,'' \emph{J. Speech Lang. Hear. Res}, 2009.

\bibitem{kourkounakis2020b}
T.~Kourkounakis, A.~Hajavi, and A.~Etemad, ``Fluentnet: End-to-end detection of
  speech disfluency with deep learning,'' in \emph{arXiv}, 2020.

\bibitem{riad2020identification}
R.~Riad, A.~Bachoud-L{\'e}vi, F.~Rudzicz, and E.~Dupoux, ``Identification of
  primary and collateral tracks in stuttered speech,'' in \emph{LREC}.\hskip
  1em plus 0.5em minus 0.4em\relax European Language Resources Association,
  2020.

\bibitem{DasNSRE2020}
A.~Das, J.~Mock, H.~Chacon, F.~Irani, E.~Golob, and P.~Najafirad, ``Stuttering
  speech disfluency prediction using explainable attribution vectors of facial
  muscle movements,'' in \emph{arXiv}, 2020.

\bibitem{Mendelev2020}
V.~Mendelev, T.~Raissi, G.~Camporese, and M.~Giollo, ``Improved robustness to
  disfluencies in rnn-transducer based speech recognition,'' in \emph{arXiv},
  2020.

\bibitem{alharbi2018lightly}
S.~Alharbi, M.~Hasan, A.~J. Simons, S.~Brumfitt, and P.~Green, ``A lightly
  supervised approach to detect stuttering in children's speech,'' in
  \emph{Interspeech 2018}.\hskip 1em plus 0.5em minus 0.4em\relax ISCA, 2018,
  pp. 3433--3437.

\bibitem{AlharbiWOCCI17}
S.~Alharbi, A.~Simons, S.~Brumfitt, and P.~Green, ``Automatic recognition of
  children's read speech for stuttering application,'' \emph{International
  Workshop on Child Computer Interaction}, vol. WOCCI, 2017.

\bibitem{HeemanInterSpeech16}
P.~Heeman, R.~Lunsford, A.~McMillin, and J.~Yaruss, ``Using clinician
  annotations to improve automatic speech recognition of stuttered speech,'' in
  \emph{Interspeech}, 2016.

\bibitem{andrews1964syndrome}
G.~Andrews and M.~Harris, ``The syndrome of stuttering.'' \emph{Spastics
  Society Medical Education}, 1964.

\bibitem{gondala2021error}
S.~Gondala, L.~Verwimp, E.~Pusateri, M.~Tsagkias, and C.~Van~Gysel,
  ``Error-driven pruning of language models for virtual assistants,''
  \emph{arXiv preprint arXiv:2102.07219}, 2021.

\bibitem{huang2020sndcnn}
Z.~Huang, T.~Ng, L.~Liu, H.~Mason, X.~Zhuang, and D.~Liu, ``Sndcnn:
  Self-normalizing deep cnns with scaled exponential linear units for speech
  recognition,'' in \emph{ICASSP 2020-2020 IEEE International Conference on
  Acoustics, Speech and Signal Processing (ICASSP)}.\hskip 1em plus 0.5em minus
  0.4em\relax IEEE, 2020, pp. 6854--6858.

\bibitem{chen2019active}
X.~C. Chen, A.~Sagar, J.~T. Kao, T.~Y. Li, C.~Klein, S.~Pulman, A.~Garg, and
  J.~D. Williams, ``Active learning for domain classification in a commercial
  spoken personal assistant,'' \emph{arXiv preprint arXiv:1908.11404}, 2019.

\bibitem{ranantha2020intent}
R.~Anantha, S.~Chappidi, and W.~Dawoodi, ``Learning to rank intents in voice
  assistants,'' \emph{arXiv:2005.0119v2}, 2020.

\end{thebibliography}

% \begin{thebibliography}{9}
% \bibitem[1]{Davis80-COP}
%   S.\ B.\ Davis and P.\ Mermelstein,
%   ``Comparison of parametric representation for monosyllabic word recognition in continuously spoken sentences,''
%   \textit{IEEE Transactions on Acoustics, Speech and Signal Processing}, vol.~28, no.~4, pp.~357--366, 1980.
% \bibitem[2]{Rabiner89-ATO}
%   L.\ R.\ Rabiner,
%   ``A tutorial on hidden Markov models and selected applications in speech recognition,''
%   \textit{Proceedings of the IEEE}, vol.~77, no.~2, pp.~257-286, 1989.
% \bibitem[3]{Hastie09-TEO}
%   T.\ Hastie, R.\ Tibshirani, and J.\ Friedman,
%   \textit{The Elements of Statistical Learning -- Data Mining, Inference, and Prediction}.
%   New York: Springer, 2009.
% \bibitem[4]{YourName17-XXX}
%   F.\ Lastname1, F.\ Lastname2, and F.\ Lastname3,
%   ``Title of your INTERSPEECH 2021 publication,''
%   in \textit{Interspeech 2021 -- 20\textsuperscript{th} Annual Conference of the International Speech Communication Association, September 15-19, Graz, Austria, Proceedings, Proceedings}, 2020, pp.~100--104.
% \end{thebibliography}

\end{document}